\documentstyle[12pt,fleqn]{article}
\begin{document}
\baselineskip= 22 truept

\newcommand{\nc}{\newcommand}
\begin{flushright}
IP/BBSR/96-05\\
MRI/PHY/96-08\\
hep-th/9601171\\
\end{flushright}
\vspace{1cm}\begin{center} 
{\large \bf  Exact Type IIB Superstring Backgrounds}
\\ \vspace{1cm} 
{\bf {\sc Supriya Kar}$^1$},
{\bf {\sc 
Alok Kumar}$^2$ } and {\bf {\sc Gautam Sengupta}$^2$$^,$$^*$}\\
\vspace{0.5cm}
$^1$Mehta Research Institute for Mathematics\\
and Mathematical Physics\\
10, Kasturba Gandhi Marg\\
Allahabad, 211 002, INDIA.\\
e-mail: supriya@mri.ernet.in\\
\bigskip

$^2$Institute of Physics\\
Bhubaneswar 751 005, INDIA\\
e-mail: kumar@iopb.ernet.in

\end{center} 
\thispagestyle{empty}
\vskip 2cm
\begin{abstract}

We obtain a family of type IIB superstring backgrounds involving
Ramond-Ramond fields in ten dimensions
starting from a heterotic string background with vanishing gauge fields.
To this end the global $SL(2,R)$ symmetry of the type IIB
equations of motion
is implemented as a solution generating transformation. Using a 
geometrical analysis we show that the type IIB backgrounds obtained are
solutions to all orders in $\alpha^{\prime}$.

\end{abstract}
\vfil
\hrule
\leftline {$^*$Short term visitor.}
\eject

\nc {\be}{\begin{equation}}
\nc {\ee}{\end{equation}}

\nc {\bea}{\begin{eqnarray}}
\nc {\eea}{\end{eqnarray}}
\nc {\gam} {\Gamma}

\nc {\pa}{\partial}

\nc {\rc}{R_{\lambda\mu\nu\kappa}=
2l_{[\lambda}\pa_{\mu ]}\pa_{[\nu}Fl_{\kappa ]}}

\nc {\rci}{R_{\lambda\mu\nu\kappa}}
\nc {\ric}{R_{\mu\nu}}

\noindent There has been a resurgence of interest in string theory
over the last one year in the context of the duality symmetries. These
symmetries have assumed a central role in establishing
a {\it web of interconnections} amongst disparate string theories in diverse
dimensions.
They include the
T-dualities \cite{tdual} which are realized perturbatively in the string loop
expansion and the conjectured non perturbative S-duality 
\cite{sdual,senrev,jhs}.
The latter relates strong and weak coupling phases and 
electric and magnetic charged states. In the quantum theory, they are
conjectured to arise as broken subgroups of continuous global
symmetries of the classical effective field theory but are discrete
gauge symmetries of the full string theory. As an example we have the
toroidal compactification of a heterotic string theory to four
dimensions \cite{senrev}.
It has been shown that there exists a continuous global 
classical $SL(2,R)$ symmetry which is
broken to the discrete S-duality group $SL(2,Z)$ in the quantum theory.
This relates electric charged states of the fundamental string spectrum 
to the magnetic charged solitons. An
analogous situation exists for type IIB superstring 
theory in ten dimensions \cite{hull,berg}.
There we have likewise a global $SL(2,R)$ symmetry of the classical
effective field theory in ten dimensions which is broken in the
quantum string theory to the discrete S-duality group $SL(2,Z)$.
However, unlike the heterotic case the S-duality for the type IIB case 
relates electric and magnetic
charged states of gauge fields arising from the two different sectors
NS-NS and R-R of the superstring spectrum.

In the perturbative approach to the background field analysis
in string theories, the well known recipe is to construct classical solutions
and consider quantum corrections in the world sheet sigma model perturbation
expansion, treating the various background fields as couplings 
\cite{sigmarev}. 
Two dimensional conformal
invariance then requires the beta functions for these couplings to vanish
leading to the background field equations of motion to various orders in
the sigma model coupling $ \alpha^{\prime}$. The higher order contributions
to the equations of motion become significant in the regime of strong fields
and gravitational singularities and may be computed perturbatively. Such an
approach is however unsuitable for type II string backgrounds. The bosonic
excitations in type II string spectrum arise from both the NS-NS and the
R-R sectors \cite{fms}. 
The sigma model coupling of the R-R backgrounds are 
complicated, involving spin fields which depend on the ghosts. This introduces
the complex process of picture changing and break the separate
superconformal invariances of the matter and the ghost sectors. Thus 
higher order $ \alpha^{\prime}$ computations for the type II case is
an extremely nontrivial exercise \cite{fms,pol}. 
There exists however certain classes of
string backgrounds, at least for the bosonic and heterotic strings,
for which the leading order equations of motion are exact
as all higher order contributions are vanishing 
\cite{tset1,berg1,tset2,tsetk,tsetrev}. A
large class of such backgrounds are ones with a {\it covariantly constant
null killing vector} \cite{tset1,tsetk}. 
These include the K-models with plane wave backgrounds
as an explicit example. For the latter backgrounds, assuming certain specific
ansatz for the form of the field strength of the antisymmetric tensor and the
dilaton, it is posible to show from purely geometrical arguments
that all higher order terms in the equations of 
motion are vanishing \cite{horo}. This obviates any requirement for a 
higher order
perturbative analysis in the world sheet sigma model expansion.
These plane waves 
may also be considered as heterotic string backgrounds with vanishing space
time fermions and gauge field \cite{berg1,horo,guv}.

In this article we obtain a class of all order ($\alpha^{\prime}$ )
type IIB backgrounds
in ten dimensions with nontrivial R-R fields but vanishing 
five form field strength. We start from an exact 
plane wave background embedded in a heterotic string theory.
Utilising the global $SL(2,R)$ symmetry of the type IIB equations of motion
as a solution 
generating transformation \cite{berg}, we generate a type IIB background.
We subsequently use geometrical arguments to show
that all higher order contributions to the background field equations of
motion are identically zero for the type IIB case also. In this way we
are able to construct
a family of $SL(2,R)$ invariant, exact ( to all orders 
in $\alpha^{\prime}$) type IIB superstring backgrounds, completely avoiding
the complex issues of world sheet couplings of the R-R fields.
Our results acquire added significance as the plane waves are special cases
of the K-models \cite{tsetk}. Certain K-models for heterotic and type II
strings with vanishing R-R fields, known as
{\it chiral plane waves},
are connected to extremal electric and magnetic charged black holes 
\cite{tsetrev}.
These are known to be exact (in $\alpha^{\prime}$) and when
embedded in ten dimensional supergravity they preserve half of the space-time
supersymmetry.

We now begin with a description of the plane wave string background
for the heterotic case with vanishing gauge fields and 
briefly describe the geometrical arguments 
\cite{horo} to show that the leading order
equations of motion are exact. These solutions may also be trivially
embedded in type II string theories.
The general form of a plane wave metric is as follows \cite{horo}

\begin{equation}
ds^{2}=-2 dudv + dx^{i}dx_{i} + F(u,x) du^{2},
\end{equation}

\noindent where $u,v$ are light cone coordinates and $x^{i}$ (where $i=2...9$)
are the transverse coordinates. In our notation greek indices 
run over ( 0 to 9)
and the roman indices over the transverse coordinates (2 to 9).
The $v$
isometry of the metric leads to a null killing vector of the form
$l^{\mu}= (0, 1, 0,......0)$. For the metric in eqn. (1), the only non-zero
connections are 
$\gam^{i}_{uu}$, $\gam^{v}_{uu}$ and $\gam^{v}_{ui}$. Using these, it is 
possible
to show that $l^{\mu}$ is covariantly constant i.e. $D_{\mu}l^{\nu}=0$. As
the $G_{uv}$ component is a constant, we also have $D_{\mu}l_{\nu}=0$. 
The metric may then be expressed in a closed form in 
terms of the killing vector
$l^{\mu}$ as, $G_{\mu\nu}=\eta_{\mu \nu} + Fl_{\mu}l_{\nu}$,
with the inverse metric as $G^{\mu\nu}=\eta^{\mu \nu} - Fl^{\mu}l^{\nu}$.
Now the Riemann curvature tensor for this background may be
calculated and shown to be as follows,
\begin{equation}
R_{\lambda\mu\nu\kappa}=2\ l_{[\lambda}\pa_{\mu ]}\pa_{[\nu}Fl_{\kappa ]}.
\end{equation}

\nc {\half}{{1\over 2}}

Employing the killing equation for $l^{\mu}$, it is straightforward 
to show that
it is orthogonal to $R_{\lambda\mu\nu\kappa}$ on all the indices.
The Ricci tensor for the metric in eqn. (1) is $\ric = 
-\half (\pa^{2}F)l_{\mu}l_{\nu}$. 
From the form of the inverse metric we have $G^{uu}=0$ and as $F$ is 
independent
of $v$, then $\pa^{2}F=\pa_{T}^{2}F$ where $\pa_{T}^{2}=\pa^{i}\pa_{i}$
and $i$ runs over the transverse coordinates $x^{i}$. It is evident that when
$F$ is a harmonic function of the transverse coordinates, the metric (1) is
a solution of the vaccuum Einstein equation $\ric =0$.

It is now straightforward to include the antisymmetric tensor and the 
dilaton. We choose the dilaton $\phi$ to be any arbitrary function of $u$
and the 3-form field strength for the antisymmetric tensor as
\be H_{\lambda\mu\nu}=A_{ij}(u)\ l_{[\lambda}D_{\mu}x^{i}D_{\nu ]}x^{j},
\ee

\nc {\hf}{H_{\lambda\mu\nu}=
A_{ij}(u)l_{[\lambda}D_{\mu}x^{i}D_{\nu ]}x^{j}}
\nc {\hi}{H_{\lambda\mu\nu}}
The only independent tree level equation of motion for the background fields 
now is \cite{horo}
\be
\ric - {1\over 4} H_{\mu\alpha\beta}H_{\nu}^{\phantom\ \alpha\beta} - 
2D_{\mu}D_{\nu}\phi (u)=0.
\ee
As $D_{\mu}\phi$ is proportional to $l_{\mu}$ from the Killing 
equation for the
scalar dilaton field $\phi (u)$, all terms in (4) are proportional to
$l_{\mu}l_{\nu}$. Thus (4) is satisfied provided we have
\be
\pa^{i}\pa_{i}F(u,x) + {1\over {18}} A_{ij}(u)A^{ij}(u) + 
4\pa_{u}^{2}\phi (u) = 0
\ee
So we may choose $A_{ij}(u)$ and $\phi (u)$ arbitrarily and solve for 
$F(u, x)$. To satisfy eqn (5), $F(u,x)$ must be a quadratic function of
$x^i$.
There is a large class of such solutions, as we may also add
to $F(u, x)$ any solution of the homogeneous equation.

Before discussing the case of the type IIB backgrounds with nontrivial
R-R fields, we first discuss
the higher order sigma model contributions to the tree level
equation (4) for the heterotic background \cite{horo}.
From the tensor structure of the equation 
(4), it is evident that the higher order terms are all second rank 
tensors constructed from powers of the Riemann tensor $\rci$, the 3-form
field strength $\hi$, the dilaton
$\phi (u)$ and their covariant derivatives.
The first such class of higher order terms
are the ones which involve a single Riemann tensor. This has the
explicit form,
$D^{\lambda}D^{\nu}\rci$, which may be expressed in terms of the
covariant derivatives of the Ricci
tensor $\ric$ through the Bianchi identitites. As $F$ must be a quadratic
function of $x^i$ {\it cf}. eqn (5), 
$D^{\lambda}D^{\nu}\rci=0$.

From the expression of the Riemann tensor (2) in terms of the
null killing vector $l^{\mu}$, it may be shown that all second rank
tensors constructed from more than one $\rci$ involve a contraction of the
killing vector $l^{\mu}$ with itself or with $\rci$.
Thus they are all 
identically zero, as $l^{\mu}$ is orthogonal to $\rci$ and is also a 
null vector.
For the case when $l^{\mu}$ is contracted on a covariant 
derivative, using the 
fact that it is covariantly 
constant, it may be shown that 
\be
l^{\rho}D_{\rho}(D......D\rci)={\cal L}_{l}(D.......D\rci),
\ee
where ${\cal L}_{l}$ denotes a Lie derivative along the killing vector 
field and hence vanishes. Furthermore we have from (3) that $\hi l^{\nu}=0$
and employing similar arguments as above it may be shown that all
second rank tensors constructed from more than two $\hi$ and their derivatives
vanish. Similarly all terms in one $\rci$ and one or more $\hi$ or 
$D_{\mu}\phi$ are identically zero. The only remaining higher order terms 
are of the form $(D......DH)^{2}$. For this, notice that
$l_{[\lambda}D_{\mu}x^{i}D_{\nu ]}x^{j}$ is covariantly constant. So all
derivatives of $\hi$ acts on $A_{ij}(u)$ which results in more $l_{\mu}$
and hence this term vanishes. So the background fields satisfying eqn. (4) are
exact to all orders in $\alpha^{\prime}$.

After presenting the arguments that the backgrounds defined by 
equations (1), (3) and $\phi(u)$, are all order solutions 
(in $\alpha^{\prime}$), of the
heterotic string equations of motion, we now proceed to the type IIB case
with non-trivial R-R fields.
A type IIB superstring background consists of the following fields, the
string frame metric $G_{\mu \nu}$, two 3-form field strengths $\hi^{(k)}$
where $k= (1, 2)$ , two scalars $\chi$ and $\phi$
from the NS-NS and R-R sectors respectively and a 5-form field strength
$F_{\lambda \mu \nu \kappa \rho}$. The two scalars $\chi$ and $\phi$
may be combined to form a complex scalar $\lambda =\chi + i e^{-\phi}$.
So we may consider the heterotic background defined by $\phi (u)$, (1) and
(3) to be a special case of a type IIB background which has $\hi^{(2)}=0$
, $\chi=0$ and $F_{5}=0$. As shown in \cite {hull,berg}, 
type IIB strings in 
$D=10$ has a global $SL(2,R)$ symmetry at the level of the equations 
of motion \cite{berg,sch}. This acts on the type IIB 
background fields as follows :
\be
G_{\mu \nu}^{\prime}=\mid c\lambda + d \mid G_{\mu \nu},
\ee

\be 
\lambda^{\prime}= {{a \lambda + b}\over {c \lambda + d}},
\ee
and 
\be
\hi^{\prime (k)}=\Lambda \hi^{(k)},
\ee

where $\Lambda$ is an $SL(2,R)$ matrix such that
\be
\Lambda = \left( \begin{array}{cc}
d & c\\
b & a\\
\end{array}\right) ,
\ee
with $ad-bc=1$. 

Implementing the transformations (7), (8), and (9)
we generate a nontrivial type IIB background
with R-R fields starting from the trivial type IIB configuration defined
by (1), (3) and $\phi (u)$. Explicitly we have
\be 
G_{\mu \nu}^{\prime}(u, x)=f(u) G_{\mu \nu}(u, x),
\ee
where $f(u)={\big [d^{2} + c^{2}e^{-2\phi (u)}\big ]}^{\half}$
and
\be 
\lambda ^{\prime}={{iae^{-\phi} + b}\over {ice^{-\phi} + d}},
\ee
with $\lambda^{\prime}=\chi^{\prime} + i e^{-\phi^{\prime}}$ and
$\lambda=i e^{-\phi}$.
So we have the final expressions for the type IIB scalars as
\be
\chi^{\prime}(u)=
{1\over {{f(u)}^{2}}}\big [  db + ac\ e^{-2\phi}\big],
\ee
\be
\phi^{\prime}(u)=\phi (u) + 2\ ln\ f(u).
\ee
For
the 3-form field strength $H^{(k)}$, $k=1, 2$ we have
\be
\hi^{\prime (1)}=d \hi^{(1)},
\ee
and 
\be
\hi^{\prime (2)}=b \hi^{(2)}.
\ee

The new metric is given as follows
\be
ds^{2}=-2 f(u) dudv +f(u) dx^{i}dx_{i} + K (u,x) du^{2},
\ee
where $K(u, x)=f(u)F(u, x)$. A rescaling $f(u)du=dU$ of the metric 
leads
to the general form for the plane wave metric
\be
ds^{2}=-2 dUdv + {\tilde f}(U) dx^{i}dx_{i} + {\tilde K}(U,x) dU^{2}.
\ee
Dropping the tilde and rewriting $U$ as $u$ in (18) we have
\be
ds^{2}=-2 dudv + f(u)dx^{i}dx_{i} + K(u,x) du^{2}.
\ee

We now show that the type IIB background generated in equations (13)-(19), 
is also
an all order solution to the type IIB equations of motion.
To the lowest order, the type IIB equations of motion are those of
the $N=2$, $D=10$ chiral supergravity in ref. \cite{sch,berg}.
They consist of equations for the scalar fields $\phi$ and $\chi$, second
rank tensors $G_{\mu\nu}$ and $B^{(k)}_{\mu\nu}$ and a 
fifth rank antisymmetric
tensor equation for the self dual five form field strength
$F_{\mu\nu\rho\sigma\kappa}$. In studying the
all order solutions for the type IIB equations, the possible corrections
to all these equations must be considered. It is also noticed that all
the gauge fields, i.e. $B^{(k)}_{\mu\nu}$, and $D_{\mu\nu\rho\sigma}$
can appear in the higher order terms only as the
corresponding gauge invariant field strengths. As a result we consider
the higher order terms in these equations obtained from combinations
of the following quantitites; $R_{\mu\nu\rho\sigma}$, $H^{(k)}_{\mu\nu\rho}$,
$D_{\mu}\phi$, $D_{\mu}\chi$, $F_{\mu\nu\rho\sigma\kappa}$ and their 
covariant derivatives. In our case we choose $F$ to be zero,
which is obtained by setting the four-form field
$D_{\mu\nu\rho\sigma}=0$ in its definition, together with the
form of $H^{(k)}$ in eqn. (15) and (16).

As earlier, the $v$ independence leads to a null killing vector $l^{\mu}$. 
From (19) the only non-zero components of the inverse metric are
$G^{uv}=-1$, $G^{vv}=- K$ and $G^{ii}={f(u)}^{-1}$. Using these,
it may be shown
that the only non-zero components of the Christoffel 
connections are $\gam^{v}_{uu}$, $\gam^{i}_{uu}$, $\gam^{v}_{ui}$,
$\gam^{i}_{ui}$ and $\gam^{v}_{ii}$. This leads to the null killing vector
being covariantly constant i.e. $D_{\mu}l^{\nu}=0$.
We also have $D_{\mu}l_{\nu}=0$. 

To consider the background field equations of motion
we now proceed to compute the Riemann curvature tensor for the 
metric (19) of the type IIB background generated by us.
Once again it is possible to write the new metric (19)
in a closed form in terms of the null killing vector as
$G_{\mu \nu}=M_{\mu \nu} + K(u,x)l_{\mu}l_{\nu}$. Where $M_{\mu \nu}$ is
a $10 \times 10$ symmetric matrix with the only non-zero components being
$M_{ij}=f(u)\delta_{ij}$ and $M_{uv}=- 1$. 
Employing this closed form expression we
calculate the Riemann tensor to be 
\be
\rci= \rci^{(1)} (u) + \rci^{(2)} (u, x),
\ee
where we have
\be
\rci^{(1)}(u)= \rci^{(M)},
\ee
with $\rci^{(M)}$ being the Riemann tensor for the metric 
$M_{\mu \nu}$ and
\be
\rci^{(2)}(u, x)=2\ l_{[\lambda}\pa_{\mu ]}\pa_{[\nu}K\ l_{\kappa ]}.
\ee
Notice that $R^{(2)}$ in eqn. (22) is exactly of the same form as that
of the heterotic case (or the type II case with vanishing R-R
fields) in eqn. (2). However now we have an extra contribution $
R^{(1)}$ which cannot be expressed in a closed form in terms of
the killing vector $l^\mu$. This is because unlike the heterotic
case, $M_{\mu\nu}$ is a function of $u$ and not equal to the flat
diagonal metric $\eta_{\mu\nu}$. We are, however, able to
prove that the type IIB background in eqns. (13-19) is an all order
(in $\alpha^{\prime}$) solution by purely geometrical arguments.

It is apparent that the only non-zero independent component of $\rci$ is
\nc {\ru}{R_{uiui}}
$\ru$, the reason being that $M_{ij}=f(u)\delta_{ij}$ and the only non-zero
component of the killing vector $l_{\nu}$ is $l_{u}=-1$.
In the same way the only nonzero independent fully contravariant
component is $R^{vivi}$ because $G^{ij} (i\neq j)=G^{ui}=G^{vi}=0$.
Similarly we may show that the only nonzero component of the
Ricci tensor is $R_{uu}$, $R^{v}_{\phantom s u}$ and $R^{vv}$.  
From the
form of the inverse metric $G^{\mu \nu}$ it may be shown that it
is impossible to construct
any non-zero scalar from $R_{uu}$ by contraction, hence $R=0$.
The type IIB dilaton $\phi^{\prime}$ in eqn. (14) is also a
function of $u$ only, as in the heterotic case and the type
IIB 3-form field strengths $\hi^{\prime (k)}$ are
of the same form as in the heterotic case in eqn. (3). 
From (3) it may also be shown that the only non-zero
independent component of the field strength $\hi$ is $H_{uij}$.
In the
subsequent discussion we drop the primes on the type IIB fields 
defined in eqns. (13-19).

We now proceed to
present geometrical arguments to show
that the backgrounds defined in eqns. (13-19) are solutions to all
orders (in $\alpha^{\prime}$) of the type IIB equations of motion.
As earlier, we now consider
the higher order terms in the equations of motion which
are second rank tensors. These correspond to the equations for $G_{\mu\nu}$
and $B_{\mu\nu}^{(k)}$.
Terms involving just one Riemann tensor has the form
$D^{\lambda}D^{\nu}\rci$. 
Notice that, the $SL(2,R)$
transformation {\it cf.} eqn. (11) and the subsequent rescaling
in $u$ does not
affect the quadratic $x^i$ structure of $F(u,x)$. So the final $K(u, x)$ in
eqn. (19) is
also a quadratic function of $x^i$. Using these we have 
\be
D^{u}D^{u}R_{uiui} = D^{i}D^{i}R_{iuiu}
= D^{u}D^{i}R_{uiiu}
= D^{i}D^{u}R_{iuui}
= 0,            \label{DR}
\ee
which implies 
\be
D^{\lambda}D^{\nu}\rci=0.
\ee

It is apparent now that to construct second rank tensors it is required
to contract at least two indices of $\rci$ with other $R$ or derivatives
of $R$. Potentialy non-zero contributions to these terms may come from
the contractions of covariant indices $(u, i)$ and contravariant
ones $(v, i)$. Contractions on derivatives have been shown to be zero.
Contractions on other $R$ requires at least a covariant index $v$ or 
contravariant index $u$ which are unavailable. Hence we conclude that it
is impossible to construct non-zero second rank tensors from contraction of
$\rci$ and its derivatives. Hence all such higher order
contributions are vanishing.

We now focus on other higher order terms which are obtained from $D_{\mu}\chi
(u)$, $D_{\mu}\phi (u)$, $\hi^{(k)}$ and $\rci$. As earlier, from the killing
equation for the scalar fields $\chi$, $\phi$ we find that $D_{\mu}\phi$ is
proportional to $l_{\mu}$. Also, from the killing equation, 
it maybe shown that
the killing vector $l_{\mu}$ is orthogonal to $R$ in all the indices. This
shows that terms involving $D_{\mu}\chi R^{\lambda\mu\nu\kappa}$ are
identically zero. Similarly terms of the form $D^{\mu}\phi\rci$ are also zero.

For the terms obtained from more than two $\hi^{(k)}$ 
and their derivatives we have the following 
observations. As $\hi$ contains one killing vector,
all terms in more than two $H$ or its derivatives
contains more than two killing vectors $l^\mu$. As we have
only two free indices one of these must be contracted with
another $H$ or a derivative. Thus all such terms vanish as
$\hi^{(k)}l^{\lambda}=0$ and contraction on a derivative
is equivalent to a Lie derivative in the direction of the
killing vector of the tensor being considered, which is also
zero. In a similar fashion we may show that terms involving
$D\phi H$ ,$D\chi H$ and their derivatives 
are vanishing as $D_{\mu}\phi$
or $D_{\mu}\chi$ are proportional to $l_{\mu}$ which is
orthogonal to $\hi$ in all the indices. 

Let us now consider second rank tensors which may be obtained
from one or more $H$ and $R$ and their derivatives. As an
example, consider
$H_{\alpha\beta\gamma}D_{\kappa}R_{\lambda}^{\phantom s
\beta\gamma\kappa}$. This is non-zero for $\alpha=u\ or\ (i,j)$.
For $\alpha=u$, we have $(\beta , \gamma) = (i, j)$ because the only non-zero
independent component of $H$ is $H_{uij}$. However the only 
contravariant transverse indices of non-zero $R$,
available for contraction
are as described earlier, are $(i,i)$ or $(j,j)$
, hence the term is zero for $\alpha=u$. For $\alpha=(i,j)$,
$(\beta , \gamma )$ must be $(u, j)$ and must be contracted with a
contravariant index $u$ in $R$ which is unavailable. Hence the
the term is also vanishing for $\alpha=(i,j)$. Exactly similar
arguments may be used to show that terms of the generic form
$(DH)R$ as well as  
terms like ${(D...DH)}^{2}$ are vanishing.

We have therefore examined all the possible second rank tensors,
in more than two derivatives, constructed from $R$, $H^{(k)}$
and their covariant derivatives together with $D \phi$ 
and $D \chi$. To convince the reader that these 
are all the possible
forms to be examined, we once again stress that only possible
covariant tensor components are $D_u \phi$, $D_u \chi$,
$R_{uiui}$ and its covariant
derivatives with respect to $D_u$ and $D_j$, $H^{(k)}_{u ij}$ and
its covariant derivatives with respect to $D_u$.
One also has the corresponding contravariant
tensor components. 
In addition we have the identities in equation
(\ref{DR}). Using these it is possible to show that all
second-rank tensors with higher derivatives vanish. 
A similar argument shows that all possible higher 
order corrections to the
equations of motion for the scalars and to the 
self-duality condition for
$F_{\mu\nu\rho\sigma\kappa}$ which appears as 
fifth rank fully antisymmetric tensors are also zero.
Thus we have shown that the type IIB background 
obtained in eqns. (13-19) 
are exact to all orders in $\alpha^{\prime}$.

To conclude, we have obtained a class of exact 
( to all orders in $\alpha^{\prime}$)
type IIB backgrounds with R-R fields
dependent on the parameters of the
$SL(2,R)$ group. We emphasise that such an exercise in the sigma
model framework is highly nontrivial owing to the non standard forms
of the couplings for the R-R fields {\it cf.} 
\cite{fms,pol}. However this
issue could be completely bypassed through the geometrical
analysis of the higher order terms in the equations of motion. 
The plane wave backgrounds considered here are similar in nature to
a larger class of string backgrounds called K-Models, 
which have
been discussed by Tseytlin et. al. \cite{tsetk} for heterotic and type II
strings in the absence of R-R fields. These are distinguished
by the presence of a null killing isometry such that the corresponding 
killing vector is {\it covariantly constant}. It would be interesting
to show that the general type II K-models with non trivial R-R
fields are also all-order 
solutions of the
background field equations. We hope to report on this issue very soon.

\noindent {\bf Acknowledgements:} We would like to thank Koushik
Ray and Anindya Biswas for useful suggestions and interesting
discussions. S.K and G.S would like to thank Institute of
Physics, Bhubaneswar and the High Energy Theory Group there for
the warm hospitality extended to them.
\vfil\eject

\end{document}